\documentclass[aps,prb,twocolumn,superscriptaddress,letterpaper]{revtex4}
\usepackage{amsmath}
\usepackage{dcolumn}
\usepackage{epsfig}
\usepackage{graphicx}
\usepackage{latexsym}
\usepackage{amssymb}
\usepackage{color}

\begin{document}
\title{Weak antilocalization and disorder-enhanced electron interactions in crystalline GeSbTe}
\author{Nicholas P. Breznay}
\affiliation{Department of Applied Physics, Stanford University, Stanford, CA 94305, USA}
\author{Hanno Volker}
\affiliation{I. Physikalisches Institut (IA), RWTH Aachen University, 52056 Aachen, Germany}
\author{Alexander Palevski}
\affiliation{School of Physics and Astronomy, Raymond and Beverly Sackler Faculty of Exact Science, Tel-Aviv University, 69978 Tel-Aviv, Israel}
\author{Riccardo Mazzarello}
\affiliation{Institut f\"ur theoretische Festk\"orperphysik, RWTH Aachen University, 52056 Aachen, Germany}
\affiliation{JARA – Fundamentals of Future Information Technology, RWTH Aachen}
\author{Aharon Kapitulnik}
\affiliation{Department of Applied Physics, Stanford University, Stanford, CA 94305, USA} 
\affiliation{Department of Physics, Stanford University, Stanford, CA 94305, USA}
\author{Matthias Wuttig}
\affiliation{I. Physikalisches Institut (IA), RWTH Aachen University, 52056 Aachen, Germany}
\affiliation{JARA – Fundamentals of Future Information Technology, RWTH Aachen}

\date{\today}

\begin{abstract} 

Phase change materials can be reversibly switched between amorphous and crystalline states and often show strong contrast in the optical and electrical properties of these two phases.  They are now in widespread use for optical data storage, and their fast switching and a pronounced change of resistivity upon crystallization are also very attractive for nonvolatile electronic data storage.  Nevertheless there are still several open questions regarding the electronic states and charge transport in these compounds.  In this work we study electrical transport in thin metallic films of the disordered, crystalline phase change material Ge$_1$Sb$_2$Te$_4$.  We observe weak antilocalization and disorder enhanced Coulomb interaction effects at low temperatures, and separate the contributions of these two phenomena to the temperature dependence of the resistivity, Hall effect, and magnetoresistance.  Strong spin-orbit scattering causes positive magnetoresistance at all temperatures, and a careful analysis of the low-field magnetoresistance allows us to extract the temperature dependent electron dephasing rate and study other scattering phenomena.  We find electron dephasing due to inelastic electron-phonon scattering at higher temperatures, electron-electron scattering dephasing at intermediate temperatures, and a crossover to weak temperature dependence below 1 K.

\end{abstract}

\maketitle

\section{Introduction}

Phase change materials such as Ge$_2$Sb$_2$Te$_5$ or Ge$_1$Sb$_2$Te$_4$ can be rapidly and reversibly switched between the amorphous and crystalline state. This phase transformation is accompanied by a significant change of optical properties, which is exploited in rewritable optical data storage.~\cite{wuttig2007}  Phase change materials are also attractive for non-volatile electronic memories, where a pronounced change of resistance~\cite{friedrich00} and fast electrical switching~\cite{bruns09, loke12} are advantageous.  At present there are two challenges for commercial application of phase change materials in electronic memories, and both are related to their electrical transport properties.  First, the change of resistivity of the amorphous phase with time (``drift'') is a disadvantage if multi-level storage concepts are to be realized.  And second, the resistivity of the crystalline state in phase change materials is often low, so that high currents must be applied to heat the crystalline film to the melting point and subsequently bring the material to the amorphous state. To identify phase change materials with a higher resistivity in the crystalline state, an in-depth understanding of charge transport is a prerequisite.  Recent transport studies~\cite{siegrist2011} have revealed a disorder-tuned metal-insulator transition in films of crystalline GeSbTe (GST) compounds.  In this work it was shown that annealing of crystalline films to progressively higher temperatures is accompanied by a change from a negative temperature coefficient of resistivity (TCR) at low annealing temperature, to a positive TCR at higher annealing temperature.  This transition from an insulating to a metallic state is caused by a distinct increase of order upon annealing, and such an increase should have a pronounced impact on the low temperature resistance.  In addition, a prominent change of the magnetoresistance at very low temperature is expected upon tuning the order, i.e. upon annealing of the samples.  To investigate these predictions we have studied the low temperature electrical transport properties of thin, quasi 2-dimensional (2D) films.

Studies of electrical transport in low-dimensional disordered conductors have revealed several novel phenomena that can appear at low temperature, including weak localization and antilocalization quantum interference (QI) and many body disorder enhanced electron-electron Coulomb interaction (EEI) effects\cite{lee1985}.  The contributions of these two effects to the temperature dependence of the resistivity and Hall effect, as well as the magnetoresistance (MR), can be used to determine scattering mechanisms and other materials properties.~\cite{2dwlreview}  In 2D films these contributions are logarithmic in temperature and have nontrivial dependences on the magnetic field.  Magnetoresistance measurements in both perpendicular and parallel field geometries are useful for resolving these contributions, since orbital QI effects are sensitive to field orientation and can be suppressed in an applied magnetic field, while the EEI effects are isotropic and generally much less sensitive to the magnetic field.  These techniques are well established\cite{2dwlreview} but continue to be useful in the study of electrical transport in a variety of novel systems, including graphene\cite{tikhonenko2009}, unusual oxide heterointerface structures\cite{caviglia2010,wong2010}, and topological insulators \cite{he2011}.

There have been few experimental studies of low-temperature transport in phase change materials, and none investigating disorder-induced quantum effects in quasi-2D films, despite considerable theoretical and practical interest in such results. Along with the recent studies of metal-insulator transition in 100 nm thick films~\cite{siegrist2011}, there have been limited investigations of longitudinal and Hall resistance in GST materials~\cite{lee2005} and also of weak localization in the related compound GeMnTe~\cite{lim2011}.

In this article, we report the first study of disorder-induced quantum corrections, including weak antilocalization and enhanced EEI, in thin quasi-2D GST films. At high temperatures we observe a classical parabolic magnetoresistance and a large temperature independent contribution to the Drude resistivity.  This contribution comes from static disorder and decreases monotonically with increasing annealing temperature.  This large disorder gives rise to a short electronic mean free path indicative of diffusive electronic conduction.  At low temperatures ($<$ 20 K), we find a resistance minimum followed by a small upturn in the resistance that is proportional to ln(T) as the temperature continues to decrease.  The low temperature MR  is positive everywhere and shows a sharp cusp that develops below 20 K, suggesting weak antilocalization QI arising from diffusive carrier transport in the presence of strong spin-orbit scattering.  At high fields, we are able to recover the EEI contribution to both the resistivity and Hall effect.  Finally, we analyze the MR measurements using established localization theories and determine contributions to electronic scattering.  In future studies we also intend to investigate systematic changes of QI and EEI effects upon the transition from the metallic (weakly localized) to the insulating (strongly localized) state.  The data and conclusions presented in this manuscript should help to understand the relationship between structural and electronic properties~\cite{xu2012}, recently reported high temperature magnetoresistive effects~\cite{tominaga2011}, and the role of atomic vacancies in electronic properties~\cite{sun2011}.

Our paper is organized as follows.  In Sec.~\ref{sec:expt} we review the experimental parameters for our GST films, and in Sec.~\ref{sec:dim} discuss the film dimensionality.  Section~\ref{sec:logt} contains an analysis of the low temperature resistance and Hall effect anomaly resulting from QI and EEI effects.  Section~\ref{sec:mr} presents the measurements and analysis of the low temperature magnetoresistance in perpendicular and parallel field, and Sec.~\ref{sec:scat} contains a detailed discussion of the dominant scattering phenomena determined from the MR measurements.  Finally, Sec.~\ref{sec:summary} summarizes our results and suggests open questions that may be of interest for future study.

\section{Experiment and Sample Characterization}
\label{sec:expt}

\begin{figure}[!tb]
\centering
\includegraphics[width=1.0\columnwidth]{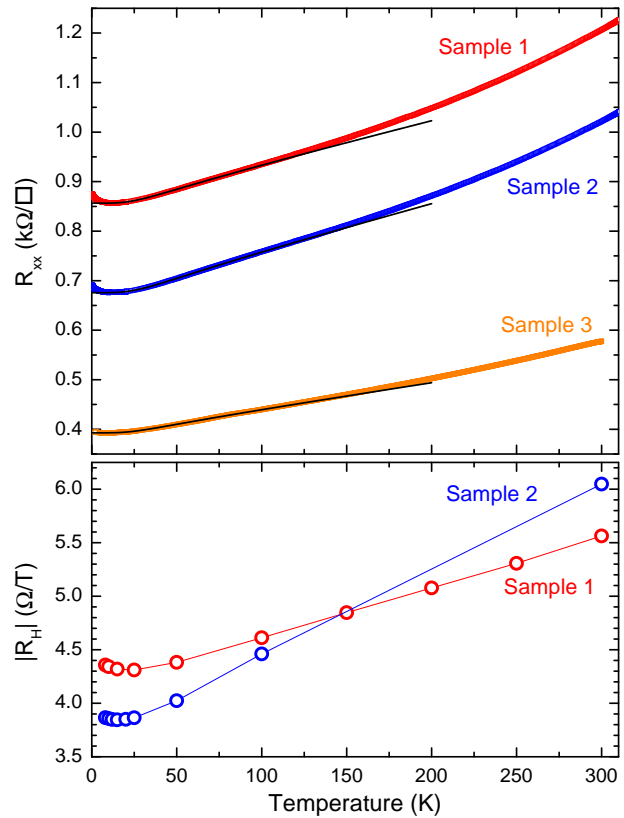}
\caption{ \footnotesize \setlength{\baselineskip}{0.8\baselineskip} Upper panel: sheet resistance $R_{xx}$ versus temperature for three GeSbTe samples showing metallic slope, large residual resistance R(0), and a low-temperature minimum.  Also shown are fits (solid black lines) to a Bloch-Gr\"uneisen form (see text). Lower panel: Hall coefficient $R_H$ versus temperature for samples 1 and 2.}
\label{fig:rvst}
\end{figure}

Films of Ge$_{1}$Sb$_{2}$Te$_{4}$ with thicknesses of 7.5 and 14 nm were deposited by sputtering onto Si substrates using stoichiometric GST targets, and capped with a 7 nm ZnS-SiO$_2$ layer (with a composition ratio of ZnS:SiO$_2$ equal to 80\%:20\%).  The film thicknesses were controlled by adjusting the sputter time as confirmed using X-ray reflectivity techniques.  Films were annealed for 30 minutes in pure Argon gas flow at temperatures of 275 C (samples 1 and 3) and 300 C (sample 2); 100 nm thick samples prepared identically were found to be in the hexagonal crystal phase.~\cite{siegrist2011}  After annealing, Hall-bar devices were patterned using conventional photolithography and Ar-ion milling techniques, with Ti/Au pads for electrical contact.  The active area of the devices was 200 $\times$ 100 $\mu$m$^2$.

\begin{table*}[!bt]
\begin{center}
\caption{Measured and calculated GST sample parameters.  The sheet resistance R$_{xx}$, carrier density $n$, mean free path $\ell_e$, and diffusion coefficient $D_{tr}$ are measured or calculated from 15 K data.  (We assume the sample 3 carrier density to be equal to sample 1.) The mobility extracted from the magnetoresistance (MR) $\mu_{MR}$ and from the Hall effect $\mu_{Hall}$ were determined using 300 K MR data. T$_{min}$ is the resistance minimum temperature, and d the film thickness.}
\begin{ruledtabular}
\begin{tabular}{c c c c c c c c c c}
Sample	& $T_{anneal}$	& $d$ &	$R_{xx}$ 				&	$n$								 		& $\ell_e$	&	$T_{min}$	&	$\mu_{Hall}$	&	$\mu_{MR}$		& $D_{tr}$	\\
				&	($^{\circ}$C)	& (nm)& (k$\Omega/\Box$)&	(10$^{20}$ cm$^{-3}$) & (nm)			&	(K)				&	(cm$^2$/V s)	&	(cm$^2$/V s)	& (cm$^2$/s)\\
\hline
1				&	275						& 7.5 &	0.86						&	1.9										& 3.7				& 14				&	35.7					&	61						& 4.05			\\
2				&	300						& 7.5	&	0.68						& 2.2 									& 4.4				& 12				&	37.5					&	73						&	4.94			\\
3				&	275						&	14.0&	0.39						&	(1.9)									&	4.4				&	9					&	58.7					&	62						&	4.73			\\
\end{tabular}
\end{ruledtabular}
\end{center}
\label{tab:paramtab}
\end{table*}

The longitudinal and Hall resistances were measured in several Quantum Design PPMS cryostats in applied magnetic fields of up to 9 T using standard four-point dc and low frequency lock-in techniques.  Three completely independent experimental set-ups were used and all gave identical results.  Care was taken to ensure that resistances were measured in the ohmic regime, especially at low temperatures ($<$ 1 K).  Over 40 devices from the three films were fabricated and characterized at room temperature, and several for each film were further studied at low temperatures.  All devices from the same film showed qualitatively identical behavior and only minor variation in their resistivity, Hall coefficient, and other parameters.  All experimental data and formulas below report resistances $R_{ij}$ and conductances $\sigma_{ij}$ in 2D (sheet) values.

Figure~\ref{fig:rvst} shows the measured sheet resistance versus temperature for all samples, which show a positive temperature coefficient of resistance and weakly metallic behavior with a large zero-temperature residual resistance R(0).  The decrease in the resistance between 300 and 20 K is consistent with a reduction in electron-phonon scattering; Fig.~\ref{fig:rvst} shows fits to the Bloch-Gr\"uneisen expression\cite{ziman} for the phonon contribution to electronic scattering, $R(T)/R_{min} -1 \sim (T/\theta_D)^5$ at the lowest temperature.  We find a Debye temperature $\theta_D \approx 140-150$ K, consistent with independent measurements on comparable films~\cite{zalden11}, and extract a value for the electron phonon coupling constant $\lambda_{ep} \sim 0.1$.

The measured Hall resistance $R_{xy}$ is linear in the applied magnetic field and is consistent with p-type charge carriers; we extract the Hall coefficient from high-field (B $>$ 2 T) measurements of $R_{xy}$.  The lower portion of Fig.~\ref{fig:rvst} shows the Hall coefficient $R_H = R_{xy}/B$ as a function of temperature for samples 1 and 2, which like the longitudinal resistance shows a low temperature minimum.  The measured Hall coefficient at 15 K yields a carrier density $n \sim 2 \times 10^{20} cm^{-3}$ via $|R_{H}| = \frac{1}{n e t}$ and a Hall angle Tan $\theta_H = R_{xy} / R_{xx} \approx$ 0.005 at a field of 1 T. 

The longitudinal conductance in the presence of a magnetic field $\sigma_{xx}(B)$ can be calculated from $R_{xx}$ and $R_{xy}$ through
	\begin{equation}
	\sigma_{xx}(B) = \frac{R_{xx}}{R^2_{xx} + R^2_{xy}} = \frac{1}{R_{xx}}\left(1+\tan(\theta_H)^2\right)^{-1}
	\end{equation}
Since $\tan(\theta_H) \ll 1$, the conductance $\sigma_{xx} \approx 1/R_{xx}$, and the magnetoconductance $\Delta\sigma_{xx}(B) = \sigma_{xx}(B) - \sigma_{xx}(0) \approx 1 / R_{xx}(B) - 1/R_{xx}(0)$.  Below we plot and analyze the negative magnetoconductance, $-\Delta\sigma_{xx}(B)$, which has the same sign as the MR: $-\Delta\sigma_{xx}(B) \approx \Delta R_{xx}(B)/R_{xx}(0)^2$.

\begin{figure}[ht]
\centering
\includegraphics[width=1.0\columnwidth]{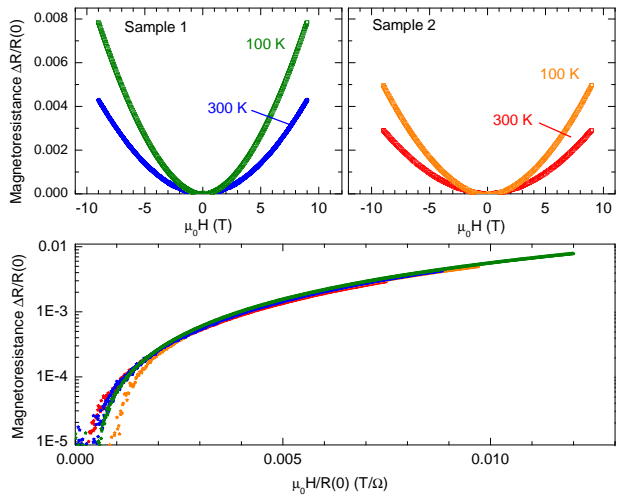}
\caption{ \footnotesize \setlength{\baselineskip}{0.8\baselineskip} Top panels: magnetoresistance for samples 1 and 2 at 300 and 100 K. Lower panel: Kohler diagram showing scaling of all the magnetoresistance datasets in the upper panels.}
\label{fig:classmr}
\end{figure}

Now let us consider the normal state magnetoresistance.  Figure ~\ref{fig:classmr} shows the measured MR at 300 K and 100 K for samples 1 and 2, defined as
	\begin{equation}
	\frac{\Delta R}{R(0)} = \frac{R_{xx}(B) - R_{xx}(0)}{R_{xx}(0)}.
	\end{equation}
At these temperatures, the MR is small, positive, and parabolic.  Kohler's rule\cite{ziman} states that in a classical metal with one dominant scattering time, the magnetoresistance $\frac{\Delta R}{R(0)}$ should be a universal function of the quantity $\omega_c \tau$, where $\tau$ is the transport scattering time, the cyclotron frequency $\omega_c \equiv e B/m^*$, and $m^*$ is the effective mass.  At sufficiently low fields
	\begin{equation}
	\frac{\Delta R}{R(0)} \sim \left( \omega_c \tau \right)^2,
	\end{equation}
consistent with the data presented in Fig.~\ref{fig:classmr}. From parabolic fits to the 300 K curve for sample 1 we derive a transport scattering time of $\tau \approx 1.4 \times 10^{-14}$ s and a mobility $\mu \approx$ 61 cm$^2$/(V s), comparable to the calculated Hall mobility $\mu_{Hall} \approx$ 36 cm$^2$/(V s) and somewhat larger than that achieved in thick (100 nm) films annealed to the same temperatures~\cite{siegrist2011}.  For free electrons $\omega_c\tau = B R_H / R_{xx}$ and so the normal state MR should be a universal function of $B/R_{xx}$; such scaling for samples 1 and 2 is shown in the lower panel of Fig.~\ref{fig:classmr}, consistent with a free-electron, single scattering time picture of the classical electrical transport from 100 - 300 K.  In our consideration of the low temperature MR below, this classical contribution was also apparent at the highest fields and was subtracted before proceeding with quantitative analysis.

Based on the measured resistivity, Hall coefficient, and magnetoresistance, we calculated materials parameters including the mean free path $\ell_e$ and diffusion coefficient $D_{tr}$ as shown in Table~\ref{tab:paramtab} for each sample, assuming a free-electron like picture with effective mass m$^*$ = 0.4 m$_e$ and a valley degeneracy $N_v$ = 4.\cite{siegrist2011}

\section{Sample dimensionality}
\label{sec:dim}

Before considering the WL analysis, let us discuss the dimensionality of our GST films.  A thin film can be treated as two dimensional if the thickness d is smaller than the appropriate physical length scales.  However, due to the many interactions and phenomena that are relevant in this analysis, there is no single measure of our film dimensionality.  The electronic mean free path $\ell_e$ for the samples is $\sim$ 4 nm, less than d = 7.5 - 14 nm, and so the classical diffusive transport is three-dimensional.  For QI effects, the relevant length scale is the dephasing length $L_{\phi}$, related to the phase breaking time $\tau_{\phi}$ through $L_{\phi} = \sqrt{D \tau_\phi}$. In these samples the dephasing is dominated by inelastic scattering, both electron-phonon at high temperatures, and electron-electron scattering at lower temperatures.  The rates for both of these scattering process increase with temperature, and thus the $L_{\phi}$ decreases with temperature.  At a sufficiently low temperature there should be a crossover to 2D behavior when $L_{\phi} \sim d$, which (based on the analysis below) should occur at roughly 50-100 K.  For EEI effects, the relevant length is the thermal diffusion length, $L_T = \sqrt{D \hbar / k_B T}$, which is equal to the film thicknesses at T $\sim$ 40 K (sample 3) up to 140 K (sample 2).  Thus despite being 3D with respect to classical transport, our films can be treated as 2D for QI and EEI phenomena below $\sim$ 40 K.  We restrict our analysis to this quasi-2D limit.

The characteristic phonon wavelength $\lambda_{ph} = \hbar v_s / k_B T \sim 1$ nm at 15 K, smaller than the film thicknesses, indicating that the phonons cannot be considered two-dimensional and therefore constraining the theoretical predictions for the electron-phonon scattering rates.  Finally, the characteristic magnetic length $\ell_B = \sqrt{\hbar/4 e B}$ is equal to the 7.5 nm thickness of samples 1 and 2 at $\sim$ 3 T and equal to the sample 3 film thickness of 14 nm at $\sim$ 1 T; at higher fields the diffusive transport is no longer in the quasi-2D limit.  We now turn to an analysis of the anomalous upturn in the resistance apparent at temperatures below 20 K.

\begin{figure}[ht]
\centering
\includegraphics[width=1.0\columnwidth]{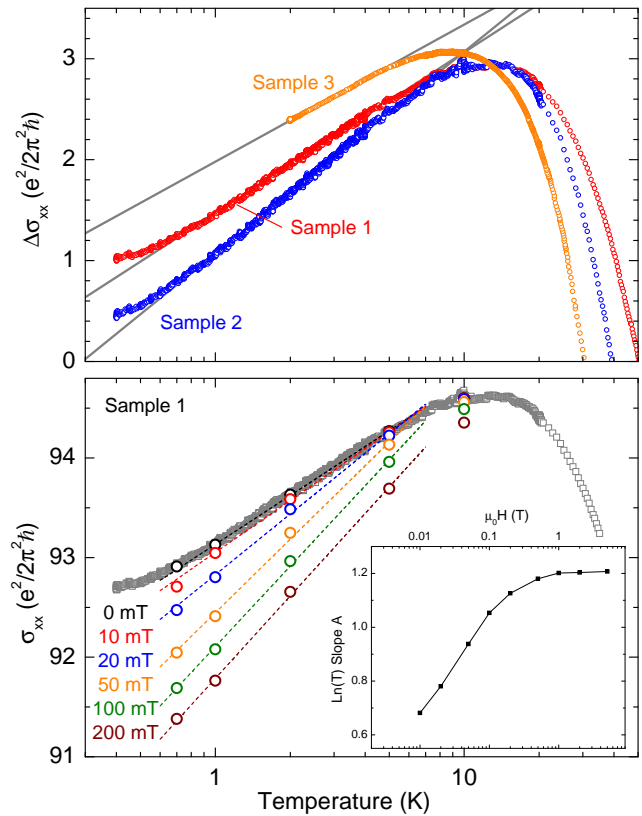}
\caption{ \footnotesize \setlength{\baselineskip}{0.8\baselineskip} Top panel: relative change in conductance versus temperature for all three GST samples along with linear fits in ln(T); the curves have been displaced vertically for clarity.  Lower panel: conductance of sample 1 in zero field and applied magnetic fields of 10, 20, 50, 100, and 200 mT. Inset: Slope of ln(T) fits for sample 1 as a function of applied magnetic field on double logarithmic scales, showing saturation at fields above $\sim$ 0.5 T.}
\label{fig:sigvst}
\end{figure}

\section{Weak Localization and ln(T) Resistance}
\label{sec:logt}

\subsection{Theoretical Background}

The low temperature electrical conductance $\sigma_{xx}$ of a disordered metallic film in the presence of weak localization QI and EEI corrections can be written as
	\begin{equation}
	\sigma_{xx}(B,T) = \sigma_0 + \Delta\sigma^{QI,EEI}(B,T)
	\label{eq:eeiqi}
	\end{equation}
where $\sigma_0 = \frac{n e^2 \tau d}{m^*}$ is the (classical) Drude conductance, and the remaining term arises from disorder-induced QI and enhanced EEI effects. These corrections have been studied extensively~\cite{lee1985} and are especially pronounced in systems of reduced dimensionality. In 2D, both effects lead to a ln(T) contribution to the conductance.

At high temperatures, when the electronic phase coherence time ($\tau_{\phi}$) is very short, the diffusive charge transport in a disordered film can be treated classically. As the temperature is decreased, electron-phonon and other inelastic scattering events (which randomize the electronic phase and thus limit $\tau_{\phi}$) are suppressed. Once $\tau_{\phi}$ becomes long relative to the elastic scattering time $\tau_e$, quantum interference effects appear.  One such effect, weak localization (WL), results in an upturn in the resistance at low temperature arising from constructive interference of self-intersecting time-reversal-symmetric trajectories.  Weak antilocalization (WAL) occurs when spin-orbit scattering is strong, and results in a decrease in the resistance.  The WL correction, $\Delta\sigma(T) \sim \ln \left[ \tau_{\phi} / \tau_e \right]$, is sensitive to the dephasing time since only trajectories that are coherent within this timescale can contribute to the interference effect.  The characteristic amplitude of these effects is $G_0 \equiv \frac{e^2}{2 \pi^2 \hbar}$, and if the inelastic scattering rate has a power law dependence on the temperature, $\tau_{\phi}^{-1} \sim T^p$, then the WL correction is proportional to ln(T)~\cite{hikami1980}
	\begin{equation}
 	\Delta\sigma^{QI}(T) = G_0 p t N_v \eta \ln \left[ T / T_0 \right].
	\end{equation}
Here $T_0$ is a reference temperature, $N_v$ is the number of degenerate valleys, $\eta$ is related to the strength of intervalley scattering, and the prefactor $N_v \eta$ is of order unity. ($N_v \eta \rightarrow 1$ when intervalley scattering is strong, and $N_v \eta \rightarrow N_v$ when intervalley scattering is weak.)  Finally, $t$ relates to the strength of spin-orbit scattering, $t$ = 1 in the weak spin-orbit scattering (WL) limit ($\tau_{so}^{-1} \ll \tau_{\phi}^{-1}$), and $t$ = $-\frac{1}{2}$ in the opposite (WAL) limit ($\tau_{so}^{-1} \gg \tau_{\phi}^{-1}$).

The EEI correction to the conductance\cite{lee1985,altshuler83a} also shows a ln(T) dependence
	\begin{equation}
	\Delta\sigma^{EEI}(T) = G_{0} \left( 1 - \frac{3}{4} F_{\sigma} \right) \ln \left[ T / T_0 \right]
	\end{equation}
where $F_{\sigma}$ is a 2D effective screening parameter and in the limit of strong spin-orbit scattering $F_{\sigma} \rightarrow 0$.~\cite{altshuler83a,mcginnis85}  Assuming that the QI and EEI effects are additive, the total zero-field contribution to the conductance is
	\begin{equation}
	\Delta\sigma(T) = G_0 A \ln \left[ T / T_0 \right]
	\label{eq:logtfinal}
	\end{equation}
where the prefactor A is given by
	\begin{equation}
	A = N_v \eta t p + 1 - \frac{3}{4} \tilde{F}_{\sigma}.
	\label{eq:adefn}
	\end{equation}
Modest applied magnetic fields suppress the WL correction but not the EEI term, so we now consider our measurements of the conductance in zero and applied fields as a function of temperature.

\subsection{Temperature dependence of the conductance}

Figure~\ref{fig:sigvst} shows the conductance versus temperature for each sample.  Also shown are linear fits in the temperature range 0.7 K to 5 K, where the data are approximately linear in ln(T), with slopes A $\sim$ 0.7.  As shown in Eq.~\ref{eq:adefn}, these slopes contain contributions from QI and EEI effects.  Measuring this slope in applied magnetic fields allows separation of the two, since the QI effect can be suppressed with a moderate field, while the EEI contribution contributes only weakly to the MR and cannot be suppressed.  The lower portion of Fig.~\ref{fig:sigvst} shows the conductance in applied fields ranging from 0 up to 200 mT for sample 1, along with linear fits to the low-temperature data (dashed lines).  The resulting logarithmic slopes for these fits (along with additional fits at higher fields, not shown on the plot) are shown in the inset.

The slope of the longitudinal conductance is $A_{Low} \sim$ 0.70 in zero field, and saturates to $A_{Hi} \sim$ 1.2 at fields of roughly 0.5 T or more.  Based on our analysis of magnetoresistance measurements (see discussion below) we know that the characteristic dephasing field $B_{\phi}$ for suppression of the QI correction is less than 0.1 T throughout this temperature range; thus 0.5 T should be sufficient to suppress the QI localization effect.  With the QI conductance correction slope $A^{QI}$ = $A_{Low}$ - $A_{Hi} \approx$ -0.5, the remainder should be due to EEI: $A^{EEI}$ = $A_{Hi} \approx$ 1.2.  The value for $A^{QI}$ is is consistent with strong intervalley scattering ($N_v \eta \rightarrow 1$), t = -$\frac{1}{2}$ (the strong spin-orbit scattering limit), and p $\approx$ 1 as expected for electron-electron scattering dominated dephasing (see discussion below).  In the limit of very strong screening and strong spin-orbit scattering $F_{\sigma} \rightarrow$ 0, the EEI contribution $A^{EEI}$ should be 1.  Our value of 1.2 indicates either that the parameter $F_{\sigma}$ is negative or that there is an additional ln(T) contribution to the conductance appearing at low temperature beyond those considered here.

Table \ref{tab:rvstanalysis} shows the collected results from the analysis of the zero-field and in-field resistance versus temperature measurements.  We comment that the observed EEI effects seen in these metallic samples showing WL behavior does not contradict the observations of Siegrist et al.~\cite{siegrist2011}, who highlighted the importance of disorder relative to electron interaction effects to explain the MIT in thicker films of this material.  Since the QI and EEI corrections are expected to contribute to $R_H$ in distinct ways, further consideration of temperature dependent Hall effect data can supplement the above analysis.

\begin{table}[h]
\caption{Weak localization analysis of resistance versus temperature data for the three samples.  The coefficient A is determined from zero-field data, and the EEI prefactor $1-\frac{3}{4}F_{\sigma}$ from data measured in applied fields larger than  $\sim$ 0.5 T.  The QI prefactor $t p N_v \eta$ is the difference between these values.}
\begin{ruledtabular}
\begin{tabular}{c c c c}
Parameter																&	Sample 1	& Sample 2 	& Sample 3	\\
\hline
A																				&	0.71			&	0.69			& 0.56			\\
$A^{EEI}$ = $1-\frac{3}{4}F_{\sigma}$		& 1.22			&	1.18			& 1.15			\\
$A^{QI}$ = t p $N_v \eta$								&	-0.51			&	-0.49			& -0.59			\\

\end{tabular}
\end{ruledtabular}
\label{tab:rvstanalysis}
\end{table}

\subsection{Temperature dependence of the Hall effect}

\begin{figure}
\centering
\includegraphics[width=1.0\columnwidth]{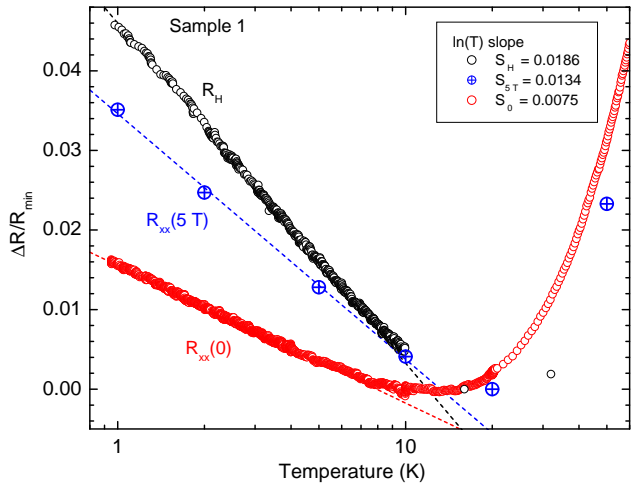}
\caption{ \footnotesize \setlength{\baselineskip}{0.8\baselineskip} Relative change in the Hall coefficient $R_{H}$ and longitudinal resistance at zero field ($R_{xx}$(0)) and 5 T ($R_{xx}$(5 T)) versus temperature for sample 1.  Also shown are linear fits (dashed lines) to the low temperature region of each proportional to ln(T).  Values for the slope S of the three linear fits are listed in the inset.}
\label{fig:HallWL}
\end{figure}

Figure~\ref{fig:HallWL} shows the Hall coefficient $R_{H}$, along with the longitudinal resistance measured in zero field, $R_{xx}$(0), and 5 T, $R_{xx}$(5 T), all as a function of temperature.  The three curves have been scaled by the resistance minimum value $R_{min}$; plotted is:
	\begin{equation}
	\frac{\Delta R_i}{R_{i,min}} \equiv \frac{R_i(T) - R_{i,min}}{R_{i,min}}
	\end{equation}
where $R_i$ is either $R_{xx}(0)$, $R_{xx}(5 T)$, or $R_{H}$.  The behavior of $R_H$ is similar to that analyzed for $R_{xx}$ in the previous section; $R_H$ shows a minimum between 10 and 20 K followed by an upturn that is linear in ln(T).  Also shown in Fig.~\ref{fig:HallWL} are linear fits to the low temperature ln(T) portion of each dataset; the slopes $S_{i}$ are listed in the inset.

Both the QI and EEI localization effects~\cite{halltheories} contribute to the diagonal conductance $\sigma_{xx}$ as described in Eq.~\ref{eq:eeiqi}, while only the QI effect contributes to the Hall conductance $\sigma_{xy}$
	\begin{equation}
	\sigma_{xy} = \sigma^D_{xy}+\Delta\sigma_{xy}^{QI}
	\end{equation}
where here $\sigma^D_{xy}$ is the classical Drude contribution to the Hall effect.  When inverting the conductivity tensor $\sigma_{ij}$, if only QI effects are present, then (to lowest order) there is no correction to $R_H$, while if only EEI effects are present, then the correction to the Hall coefficient $R_H$ is twice that to the longitudinal resistance
	\begin{equation}
	\left( \frac{\Delta R_H}{R_H}  \right)^{EEI} = 2 \left( \frac{\Delta R_{xx}}{R_{xx}}  \right).
	\label{eq:EEIonly}
	\end{equation}
Finally, if both QI and EEI corrections are present, then the Hall coefficient correction will include both QI and EEI terms\cite{mcginnis85}
	\begin{equation}
	\left( \frac{\Delta R_H}{R_H}  \right) = \frac{2 \Delta\sigma^{EEI}}{\Delta\sigma^{QI} + \Delta\sigma^{EEI}} \left( \frac{\Delta R_{xx}}{R_{xx}}  \right).
	\label{eq:RHcorr}
	\end{equation}

Since our measurements of the Hall coefficient were performed at magnetic fields much larger than the dephasing field ($B$ $\gg$ $B_{\phi}$), we would expect this to suppress the QI term in Eq.~\ref{eq:RHcorr} above, so that the correction to the Hall coefficient ($\Delta R_H / R_H$) should be 2 times the $R_{xx}$ correction as in Eq.~\ref{eq:EEIonly}.  The measured ratio ($\Delta R_H / R_H$) / ($\Delta R_{xx}(5 T) / R_{xx}(5 T)$) = $S_H$ / $S_{5 T}$ is 1.4, close to the predicted value of 2 and confirming the presence of EEI corrections at high field.  Studies of 2D electron systems in Si~\cite{bishop81uren80} and InZnO films~\cite{shinozaki07} observed ratios in the range 1-2, recovering the theoretical prediction of 2 in the $R_{xx} \rightarrow$~0 limit.

In the low-field limit we expect from Eq.~\ref{eq:RHcorr} to observe ($\Delta R_H / R_H$) / ($\Delta R_{xx}(0) / R_{xx}(0)$) = 2 $S_{5 T}$/$S_{0}$ = 3.6.  The measured ratio $S_{H}$/$S_{0}$ = 2.5 is comparable to this value, given that the Hall coefficient measurement was performed at high field.  While it is difficult to obtain precise Hall coefficient measurements in the zero-field limit, we did observe a qualitative increase in the slope $S_H$ as $B \rightarrow$~0.  This increase would give a larger value of $S_{H}$/$S_{0}$, consistent with the theoretical prediction of Eq.~\ref{eq:RHcorr}.  Additional study of the field and disorder dependence of the Hall coefficient at low temperatureshould further inform this analysis.  Having resolved QI and EEI contributions to the resistance and Hall coefficient at low T, let us now consider the WL contributions to the magnetoresistance, with the final goal of understanding the scattering mechanisms that govern the electronic dephasing processes in these materials.

\section{Magnetoresistance Analysis}
\label{sec:mr}

\begin{figure}
\centering
\includegraphics[width=1.0\columnwidth]{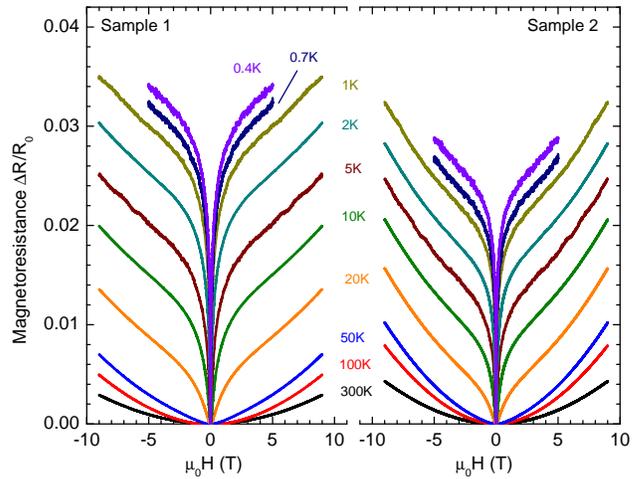}
\caption{ \footnotesize \setlength{\baselineskip}{0.8\baselineskip} Magnetoresistance for sample 1 (left) and 2 (right) from 300 to 0.4 K; temperatures are indicated for each curve.}
\label{fig:lowtmr}
\end{figure}

\subsection{Perpendicular field}
Figure~\ref{fig:lowtmr} shows the measured MR for samples 1 and 2 at low temperatures, from 300 K to 0.4 K. In this temperature region, the Kohler rule scaling of the MR breaks down, and a pronounced cusp near zero field appears.  Perpendicular magnetic fields suppress QI effects by destroying the phase coherence of time-reversal-symmetric paths.  In the case of WAL this results in a positive MR, consistent with the data in Fig.~\ref{fig:lowtmr} and indicating that we must consider spin-dependent scattering effects.  The temperature and field dependent QI correction to the conductance, in the presence of spin-orbit and spin-flip scattering, was first described by Hikami, Larkin, and Nagaoka~\cite{hikami1980} (HLN)
	\begin{equation}
	\sigma_{QI}(T,B) = - G_{0} \alpha \left[ \Psi\left(\frac{B_1}{B}\right) - \frac{3}{2} \Psi\left(\frac{B_2}{B}\right) + \frac{1}{2} \Psi\left(\frac{B_3}{B} \right)\right]
	\label{eq:hlnmr}
	\end{equation}
where $\Psi(x) = \psi(1/2 + x)$, $\psi(x)$ is the digamma function, $\alpha$ is a constant of order 1, and B$_{1,2,3}$ are given by
	\begin{align}
	B_1 = B_e + B_{so} + B_s \\
	B_2 = \frac{4}{3} B_{so} + \frac{2}{3} B_{s} + B_i \\
	B_3 = 2 B_s + B_i.
	\end{align}
Here, $B_x = \frac{\hbar}{4eD \tau_x}$ is the characteristic field corresponding to spin-orbit (so), spin-flip (s), elastic (e), and inelastic (i) scattering processes. Note that the dephasing rate $\tau_{\phi}^{-1} = 2 \tau_{s}^{-1} + \tau_{i}^{-1}$, and so we identify $B_3$ above as the corresponding dephasing field, $B_{\phi} = B_3 = 2 B_s + B_i$. In the limit of strong spin-orbit and weak spin-flip scattering, $B_{so} >> B_s$, we may further simplify the above expression
	\begin{multline}
	\Delta\sigma(B) = - G_{0} \alpha \Bigl[ \psi\left(\frac{1}{2} + \frac{B_e}{B} \right) \\
	- \frac{3}{2} \psi\left(\frac{1}{2} + \frac{\frac{4}{3}B_{so} + B_{\phi}}{B} \right) + \frac{1}{2} \psi\left(\frac{1}{2} + \frac{B_{\phi}}{B} \right) \Bigr]
	\label{eq:hlnmr1}
	\end{multline}

\begin{figure}
\centering
\includegraphics[width=1.0\columnwidth]{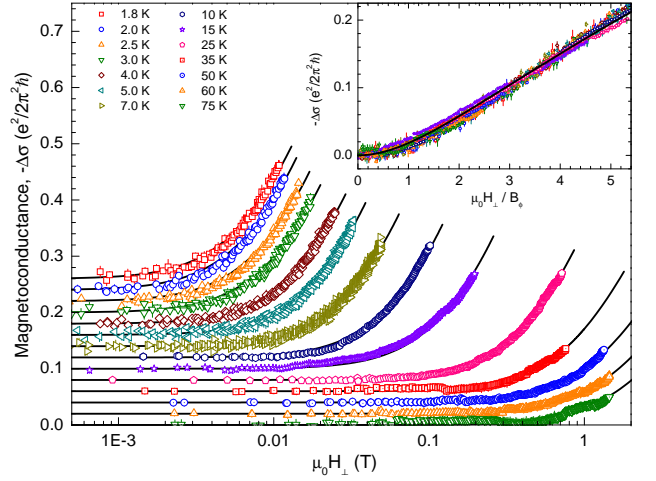}
\caption{ \footnotesize \setlength{\baselineskip}{0.8\baselineskip} Negative magnetoconductance -$\Delta\sigma_{xx}$ for sample 3 in perpendicular applied fields $\mu_0 H \sim B_{\phi}$ at temperatures from 75 - 1.8 K, plotted on a semi-logarithmic scale.  The continuous curves show single parameter fits to the theory of Hikami, Larkin, and Nagaoka~\cite{hikami1980} (HLN).  The data and curves have been offset vertically for clarity.  Inset: all magnetoconductance data from the main panel collapse to the theoretical HLN theory prediction (continuous curve) when plotted against $\mu_0 H / B_{\phi}(T)$.}
\label{fig:lowBmcFits}
\end{figure}

\begin{figure}
\centering
\includegraphics[width=1.0\columnwidth]{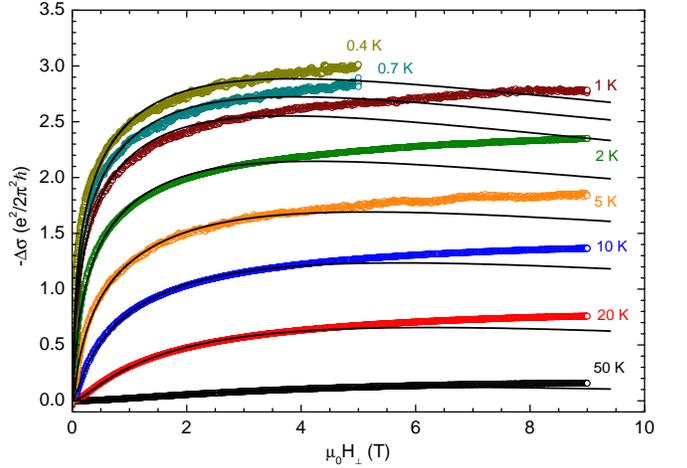}
\caption{ \footnotesize \setlength{\baselineskip}{0.8\baselineskip} Negative magnetoconductance -$\Delta\sigma$ versus applied perpendicular magnetic field for sample 1 for temperatures from 50 K to 0.4 K, along with fits (continuous lines) to the WL theory of Hikami, Larkin, and Nagaoka as described in the text.}
\label{fig:mcfits}
\end{figure}

Figure~\ref{fig:lowBmcFits} shows the negative magnetoconductance determined for sample 3 at fields comparable to $B_{\phi}$; note that the magnetic field is plotted on a logarithmic scale, and the curves are offset vertically.  The magnitude of the magnetoconductance is of order of the WL prefactor $G_{0}$.  Figure~\ref{fig:mcfits} shows the negative magnetoconductance up to fields well above  $B_{\phi}$ for sample 1.  Note that we have subtracted the (classical) parabolic contribution to the MR discussed in Sec.~\ref{sec:expt} above.  At the highest fields measured, the sign of the MR remains positive though appearing to saturate.

To reliably determine the characteristic fields appearing in Eq.~\ref{eq:hlnmr1} we first extract $B_{\phi}$ from the low-field magnetoconductance, and then $B_{so}$ and $\alpha$ using the remaining high-field data.  At low fields the classical magnetoresistance is negligible, and below $B_{\phi}$ we expect $\sim B^2$ MR behavior due to QI, since in the limit of B $\ll$ ($B_{so},B_{e}$), Eq.~\ref{eq:hlnmr} simplifies to
	\begin{equation}
	\Delta\sigma(B) = \frac{1}{48}\left( \frac{B}{B_{\phi}} \right)^2.
	\end{equation}
Representative fits used to extract $B_{\phi}$ using only the low-field portion (B $< 5 \times B_{\phi}$) of the magnetoconductance data are shown in Fig.~\ref{fig:lowBmcFits} for sample 3.  To determine the spin-orbit field we keep $B_{\phi}$ constant from the low-field analysis and considered the full magnetic field range measured.  The fits shown in Fig.~\ref{fig:mcfits} have only two free parameters - the spin-orbit field $B_{so}$ and the prefactor $\alpha$ appearing in Eq.~\ref{eq:hlnmr1}.  (The elastic scattering field $B_{e}$ was fixed based on the estimated transport scattering time $\tau \approx$ 10$^{-14}$ s, yielding $B_{e} \sim $ 10 T.)  Deviations appear at fields of above $\sim$ 3 T for this sample, however, this is already $B_{elastic}/3$ and therefore outside the region of validity for the perturbative (and diffusive-limit) HLN theory.  In addition, above this field scale the film is no longer strictly 2D, since the magnetic length $\ell_B < d$.  The parameters $B_{\phi}$ and $B_{so}$ from these fits are discussed below.  High-field deviations from the HLN result are often attributed to spin-splitting or EEI contributions to the magnetoconductance.  However, these effects should be isotropic and we observe no such phenomena appearing in the parallel field data considered next.

\subsection{Parallel fields}
When only considering orbital QI effects, a parallel magnetic field should not yield any MR with a strictly 2D film. However, as first described by Al'tshuler and Aronov~\cite{altshuler81a} (AA), a nonzero film thickness will allow for electron diffusion perpendicular to a field directed parallel to the film, and therefore lead to similar suppression of QI phenomena as in the perpendicular field case. According to AA, the resulting magnetoconductance is
	\begin{equation}
	\Delta\sigma(B_{//}) = G_{0} \ln\left[1 + \frac{B_{//}^2}{B_3 B_d} \right],
	\end{equation}
where $B_d = \frac{12 \hbar}{ed^2}$. When including spin-flip and spin-orbit scattering, this expression becomes~\cite{gershenson82gz,rosenbaum85}
	\begin{equation}
	\Delta\sigma(B_{//}) = G_{0} \left( \frac{3}{2} \ln \left[ 1 + \frac{B_{//}^2}{B_2 B_d} \right] - \frac{1}{2} \ln \left[ 1 + \frac{B_{//}^2}{B_3 B_d} \right] \right)
	\label{eq:mrpara}
	\end{equation}

\begin{figure}
\centering
\includegraphics[width=1.0\columnwidth]{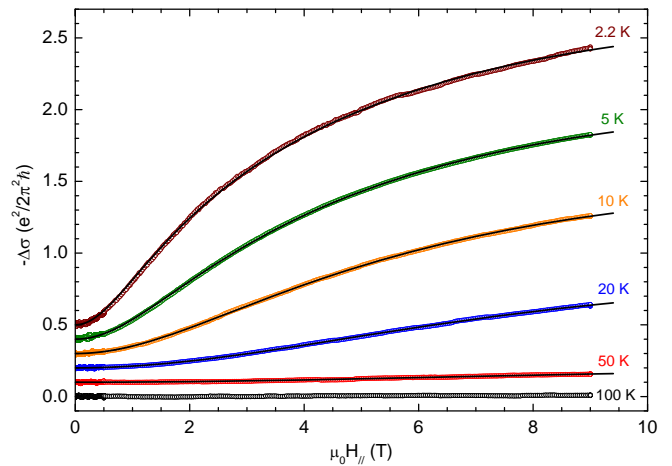}
\caption{ \footnotesize \setlength{\baselineskip}{0.8\baselineskip} Parallel field negative magnetoconductance -$\Delta\sigma_{xx}$ for sample 1, along with fits (continuous lines) to the theory of Al'tshuler and Aronov.  The curves have been vertically offset for clarity.}
\label{fig:mrpara}
\end{figure}

The magnetoconductance in parallel fields for sample 1 was also measured for a similar set of temperatures and magnetic fields; these data are plotted in Fig.~\ref{fig:mrpara}.  If the measured parallel-field magnetoresistance arose from a small out-of-plane field appearing due to misalignment of the field orientation, then the data could be simply scaled to the perpendicular field results, but no such scaling is possible.  While we cannot rule out the presence of a small out-of-plane contribution, the size of the measured magnetoconductance and lack of scaling indicates that this is not a significant contribution.  Figure~\ref{fig:mrpara} also shows fits to the theory for QI contributions to the magnetoconductance in parallel field, Eq.~\ref{eq:mrpara}, with $B_{\phi}$ and $B_{so}$ as free parameters.  The resulting values for the dephasing field $B_{\phi}$ are consistent with the perpendicular field analysis and are discussed below.  In addition, the excellent agreement up to the highest magnetic fields suggests that isotropic effects, such as arising from EEI or Zeeman splitting, may be ignored in this field range.

\begin{figure}
\centering
\includegraphics[width=1.0\columnwidth]{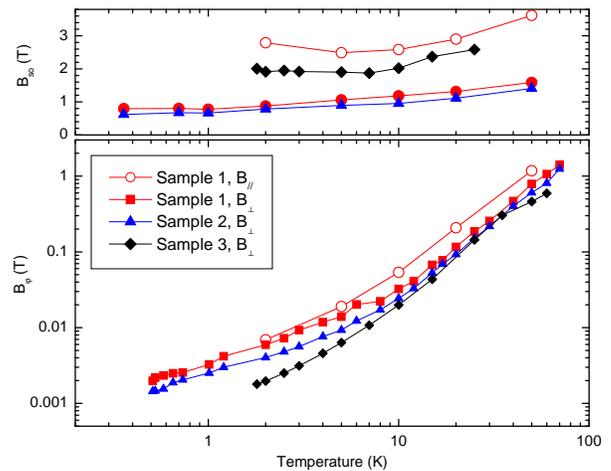}
\caption{ \footnotesize \setlength{\baselineskip}{0.8\baselineskip} Characteristic dephasing $B_{\phi}$ and spin-orbit scattering $B_{so}$ fields, extracted from fits to the magnetoconductance data, for all three samples in perpendicular (closed symbols) and parallel (open symbols) fields as a function of temperature.  The upper panel shows $B_{so}$ on a linear scale, while the lower panel shows the strongly temperature dependent $B_{\phi}$ on a logarithmic scale.}
\label{fig:params}
\end{figure}

Figure~\ref{fig:params} shows the characteristic fields associated with dephasing and spin-orbit scattering, extracted from fits to the 2D WL theory as described above.  The dephasing field $B_{\phi}$ plotted on the lower panel shows reasonable agreement between the perpendicular and parallel field measurements for sample 1, and increases strongly with increasing temperature for all three samples.  This strong temperature dependence is expected, since the field is proportional to the inelastic scattering rate which itself should increase as the temperature increases due to stronger electron-phonon and electron-electron scattering effects.  We analyze this quantitatively in the next section.  The spin-orbit scattering fields are roughly temperature independent, particularly in the zero temperature limit where $B_{so} >> B_{\phi}$, and appear to be consistent with established theory for surface scattering in the presence of high-Z atoms.

Studies of interfacial spin-orbit scattering\cite{spinorbit} found that when surface scattering is the dominant elastic mechanism, $B_{so}$ can be expressed relative to the elastic scattering field $B_e$ as $B_{so} = (\alpha_{fs} Z)^4 B_e$, where Z is the atomic number and $\alpha_{fs}$ is the fine structure constant.  Both antimony and tellurium have large Z; assuming stoichiometric Ge$_1$Sb$_2$Te$_4$ we estimate an average atomic number $\bar{Z} \sim 49$ and therefore with $B_e \approx 10$ T would expect $B_{so} \sim 0.16$ T, somewhat smaller than the $B_{so} \sim$ 0.5-2 T extracted above for the three samples.  However, the scaling of the spin-orbit times for samples 1 and 2 (both 7.5 nm thick) yields $\tau_{so, sample 1} / \tau_{so, sample 2} \sim 1.19$, comparable to the ratio of elastic scattering times $\tau_{e, sample 1} / \tau_{e, sample 2} \sim 1.13$ and  consistent with the above models.~\cite{spinorbit}  

The spin-orbit scattering rates are anisotropic and thickness-dependent, with $B^{\bot}_{so} \sim$ 0.8 T, and $B^{||}_{so} \sim$ 2.6 T for sample 1, and sample 3 (which is $\sim 2$ times thicker than 1 and 2) having $B^{\bot}_{so} \sim$ 1.94 T.  Similar studies of WAL in strong spin-orbit materials such as Bi~\cite{sangiao2011} find no such anisotropy in $B_{so}$, and in contrast to our results that the spin-orbit scattering rate decreases with increasing thickness.  Further study of high-field MR in both parallel and perpendicular field may illuminate this discrepancy and inform ongoing study of spin-orbit coupling in phase change material compounds and in related topological insulator materials such as Sb$_2$Te$_3$.

\section{Scattering rates}
\label{sec:scat}

Now let us consider the contributions to inelastic scattering in our samples, which we may compare with the characteristic dephasing rates $\tau_{\phi}^{-1}$ extracted using the magnetoresistance analysis above. In addition to temperature-independent dephasing due to spin-flip scattering from magnetic impurities or other extrinsic effects\cite{eshkol06}, the two temperature-dependent contributors to the dephasing are inelastic electron-phonon (e-p) and electron-electron (e-e) scattering. At sufficiently low temperatures, e-e scattering dephasing should dominate, since the e-p rate has a stronger temperature dependence\cite{lin2002b}. 

\subsection{Electron-electron scattering}

Al'tshuler, Aronov, and Khmelnitski (AAK)~\cite{altshuler1982} and other works\cite{lin2002b} found for the electron-electron scattering rate $\tau_{i,e-e}^{-1}$ in the 2D limit ($L_T >$ d), the following expression holds,
	\begin{equation}
	\tau_{i,(e-e)}^{-1} = \pi G_0 R_{xx} \frac{k_B T}{\hbar} \ln{G},
	\label{eq:aak}
	\end{equation}
where $R_{xx}$ is the sheet resistance of the film, and the quantity G = $\frac{\pi \hbar}{e^2 R_{xx}}$.  (We ignore an additional contribution~\cite{narozhny2002} to $\tau_{i,(e-e)}^{-1}$ that is proportional to $T^2$, since it is smaller by a factor of $k_B T / E_F \sim 10^{-3}$.)  At a temperature of 10 K, $\tau_{i,(e-e)}^{-1} \approx 10^{11}$ s$^{-1}$, or comparable to the e-p dephasing rate (calculated below) at this temperature.  At lower temperatures the dephasing field $B_{\phi}$ scales with $R_{xx}^2$ for samples 1 and 2; this is expected since both the dephasing rate and the inverse diffusion coefficient should be proportional to $R_{xx}$.  We therefore use the above theory to estimate the diffusion coefficient, since the only parameters in Eq.~\ref{eq:aak} are the film sheet resistance and fundamental constants.  For sample 1, we obtain 7.2 cm$^2/s$, comparable to the 4.05 cm$^2/s$ estimated from the Drude scattering time.  The temperature dependence of the electron-electron dephasing rate is plotted in Fig.~\ref{fig:scatrates} as a broken line, and is the most significant contribution to the total rate at intermediate temperatures.  This is consistent with the analysis of the resistance versus temperature data in Sec. III above, which showed dephasing linear in temperature between 2 and 10 K and a prefactor consistent with p = 1.  Based on this consisten picture, we use the values for the diffusion coefficient extracted using the AAK theory $D_{AAK}$ in the remainder of our analysis.

\subsection{Electron-phonon scattering}
\begin{figure}
\centering
\includegraphics[width=1.0\columnwidth]{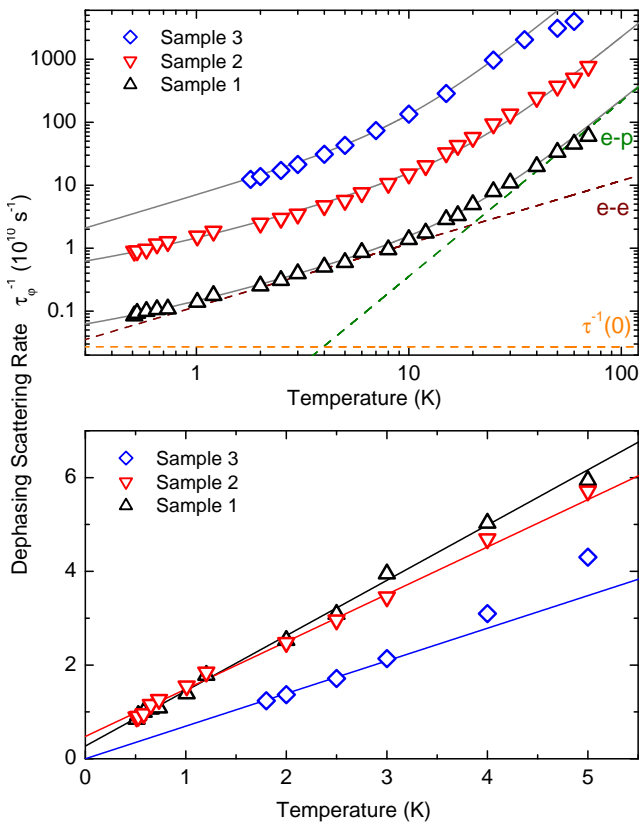}
\caption{ \footnotesize \setlength{\baselineskip}{0.8\baselineskip} Upper panel: dephasing scattering rate versus temperature on logarithmic scales for all three samples.  The solid curves show fits to a theoretical model including e-p and e-e scattering and a phenomenological T-independent dephasing rate as described in the text.  The data and fits for sample 2 (3) have been multiplied by a factor of 10 (100) for clarity.  The dotted curves show the separate contributions for sample 1 from T-independent dephasing, T-linear electron-electron scattering, and T$^{2.6}$ electron-phonon scattering. Lower panel: low-temperature portion of the same data, plotted on linear axes. The continuous curves show the best-fit contributions from e-e and T-independent dephasing terms.}
\label{fig:scatrates}
\end{figure}

There have been extensive measurements and analysis of e-p scattering contributions to dephasing in metallic and semiconducting thin films.~\cite{lin2002b} In particular, various theories have proposed, and some experiments have observed, T$^2$, T$^3$, and T$^4$ power law dependence of the scattering rate. For example, Lawrence and Meador~\cite{lawrence78} describe the electron-phonon inelastic scattering rate $\tau_{i,e-p}^{-1}$ in a dirty 2D film 
	\begin{equation}
	\tau_{i,e-p}^{-1} = 14 \pi \zeta(3) \lambda_{ep} \omega_D \left( \frac{T}{\Theta_D} \right)^3
	\end{equation}
where $\zeta$(3) is the Riemann zeta function, and $\omega_D$ and $\Theta_D$ are the Debye frequency and temperature.  With $\Theta_D$ = 150 K and $\lambda_{ep}$ = 0.1, at 10 K $\tau_{i,(e-p)}^{-1} \approx 10^{11}$ s$^{-1}$, comparable to the dephasing rate in Fig.~\ref{fig:scatrates}.  Other theories predict power laws in temperature with exponents between 2-4 for electron-phonon scattering in a 2D film.

In addition to the above T-linear and T$^x$ (with x $\sim$ 2 - 4) contributions to the electron dephasing rate, spin-flip scattering from magnetic impurities (or several other intrinsic or extrinsic processes, see discussion in e.g.~\cite{imry}) may lead to a temperature-independent contribution to the dephasing.  We therefore fit the measured dephasing rates, plotted in Fig.~\ref{fig:scatrates}, to the following form:
	\begin{equation}\label{eq:dephasingfit}
	\tau_{\phi}^{-1} = \tau_{\phi}^{-1}(0) + C_{ee} \cdot T + C_{ep} \cdot T^x.
	\end{equation}
The zero-temperature rate $\tau_{\phi}^{-1}(0)$, and the T-linear coefficient $C_{ee}$, were obtained from fits to the data below 5 K, while the parameters x and $C_{ep}$ for the electron-phonon contribution were obtained using only the high temperature (T $>$ 10 K) data.  The resulting coefficients, along with other analysis results, are collected in Table~\ref{tab:mcanalysis}, and the individual contributions for sample 1 are shown as dotted curves in Fig.~\ref{fig:scatrates}.

\begin{table}[bh]
\caption{Weak localization magnetoconductance analysis results. $D_{tr}$ and $D_{AAK}$ are the diffusion coefficient as determined from a free electron picture and by scaling the measured dephasing rate to the Al'tshuler-Aronov-Khmelnitski prediction (see text).  $B_e$ and $B_{so}$ are the elastic and spin-orbit scattering transport fields, and $\tau_{so}$	is the spin-orbit time.  The parameters $\tau_{\phi}^{-1}(0)$, $C_{ee}$, $C_{ep}$, and x describe the temperature dependent dephasing rate (see Eq. \ref{eq:dephasingfit} in the text.)}
\begin{ruledtabular}
\begin{tabular}{c c c c c}
Parameter								&	(units)									&	Sample 1	& Sample 2	& Sample 3 	\\
\hline
$D_{tr}$								&	(cm$^2$/s)							&	4.05			&	4.94			&	4.73			\\
$D_{AAK}$								&	(cm$^2$/s)							&	7.0				&	10.1			&	11.1			\\
$B_e$										&	(T)											&	20.0			&	12.6			&	11.1			\\
$B^{\bot}_{so}$					&	(T)											&	0.81			&	0.69			&	1.94			\\
$B^{||}_{so}$						&	(T)											&	2.6				&	-					&	-					\\
$\tau_{so}$							&	(10$^{-13}$ s)					&	2.9				&	2.36			&	0.76			\\
$\tau_{\phi}^{-1}(0)$		&	(10$^9$ s$^{-1}$)				&	2.7				&	4.8				&	0					\\
$C_{ee}$								&	(10$^{10}$ s$^{-1}/K$)	&	1.18			&	1.01			& 0.70			\\
$C_{ep}$								&	(10$^7$ s$^{-1}/K^x$)		&	5.9				&	17.3			&	10.5			\\
x												&	-												&	2.78			&	2.56			&	2.78			\\
\end{tabular}
\end{ruledtabular}
\label{tab:mcanalysis}
\end{table}

\subsection{Dephasing in the $T\rightarrow~0$ limit}

The lower panel of Fig.~\ref{fig:scatrates} plots the low temperature dephasing rate data for all three samples along with the theoretical T-linear e-e dephasing and temperature independent contributions.  In an ideal metallic film e-e scattering should provide the only mechanism for electronic dephasing at low temperature, and so the rate should continue to decrease as the length scale over which interference effects occur increases.  Any cutoff in this length scale, such as that due to finite sample size or another scattering mechanism, will lead to saturation of the dephasing rate or a temperature dependence that is weaker than the AAK T-linear theory.  Such saturation or weak T-dependence has been observed at low temperatures in many studies~\cite{lin2002b}, and various mechanisms have been considered.  Typically this effect is described by a temperature-independent contribution to the dephasing rate, $\tau^{-1}_{\phi}(0)$, although the physical relevance of such a description in the limit of zero temperature may be unclear.  Saturation effects may include both intrinsic mechanisms such as noise from two-level systems~\cite{imry99fs} or spin-flip scattering, and extrinsic ones, such as an ambient magnetic field, electron heating, or microwave noise effect.~\cite{lin2002lz}  Lin, Li, and Zhong observed a systematic dependence of $\tau_{\phi}^{-1}(0)$ on D in disordered metals; based on this trend and with D $\approx$ 5 - 10 cm$^2$/s the predicted saturation rate is $\approx 2 \times 10^{11} - 2 \times 10^{12} s^{-1}$, considerably higher than $\tau_{\phi}^{-1}(0)$ $\sim 3-4 \times 10^{9} s^{-1}$.  Recent studies of Sn-doped In$_2$O$_{3-x}$ films~\cite{wu2012} report a $\tau_{\phi}^{-1}(0)$ that decreases with increasing disorder, whereas sample 3 above (which has the smallest value of $R_{xx}$) shows no sign sign of a temperature independent dephasing contribution.  The magnetic field corresponding to our $\tau_{\phi}^{-1}(0)$ is $\sim$ 10 Gauss, and we estimate that residual fields within our cryostats are much less than this value.  If the zero-temperature dephasing rate is due to spin-flip scattering, then this implies a magnetic impurity concentration of roughly 10 ppm\cite{davidov1975ro} (See also the discussion in McGinnis and Chaiken~\cite{mcginnis85}).

Finally, we comment on the departure from ln(T) behavior in our R$_{xx}$ versus T data at the lowest temperatures, visible below 0.5 K in Fig.~\ref{fig:sigvst} for samples 1 and 2.  A dephasing mechanism such as spin-flip scattering will suppress the QI correction but not affect the EEI contribution.  In such a case the slope A would become larger since the QI and EEI ln(T) contributions are of opposite sign.  However, we observe a downturn in the resistance (upturn in $\sigma_{xx}$) in the T$\rightarrow$~0 limit, suggesting that the electronic system itself is unable to cool effectively at these temperatures.  Assuming that the ln(T) behavior does remain, we scaled the above scattering rates with `effective' values for the temperature assuming continued ln(T) behavior.  While the tendency towards saturation becomes less pronounced after this analysis, the temperature dependence is still weaker than T-linear, and we conclude that insufficient cooling is not responsible for the T-independent contribution to the dephasing $\tau_{\phi}^{-1}(0)$ in the above analysis.

\section{Summary and Conclusions}
\label{sec:summary}

Our analysis has shown that even the most metallic phase change films studied here are governed by strong defect scattering and hence
qualify as `dirty metals', with a rich physics of QI and EEI effects.  Significantly, in these materials we have a simple way of tuning the disorder without changing composition.  This provides an excellent opportunity to study systematic trends as a function of increasing disorder.  By analyzing three samples with different room temperature sheet resistances we are able to study disorder and thickness dependence and anisotropy of spin-orbit scattering, strengthening of the electron-phonon scattering rate with increasing disorder, and suggestions of a sub-T-linear power law governing electronic dephasing in the low-temperature limit.  Continuing such an analysis of more disordered films towards the limit of strong localization promises new insight into disorder induced MITs.  In particular we expect the weak localization approximation to break down.  Our preliminary data analysis already reveals two surprising facets, a departure from ln(T) behavior and apparent saturation in the zero-temperature limit.

We have studied the low temperature magnetotransport properties of disordered thin films of the phase-change compound Ge$_1$Sb$_2$Te$_4$ that have been annealed to be weakly metallic. We observe clear signatures of weak antilocalization and disorder enhanced electron-electron interaction effects in the resistivity, magnetoconductance, and Hall effect measurements.  Using established WL theory, we are able to extract several important materials parameters, including characteristic spin-orbit and inelastic scattering rates.  We observe a ln(T) quantum correction in $R_{H}$ 1.4 times larger than that seen in $R_{xx}$, comparable to the factor of 2 expected when the QI effects are suppressed at high field.  The inelastic scattering is dominated by phonons at high temperatures, electron-electron scattering at low temperature, and shows a weak temperature dependence at the lowest temperatures studied.  The existence of saturation or a cutoff in the dephasing length at very low temperature has been studied extensively, and both intrinsic (spin-flip scattering due to magnetic impurities) and extrinsic (coupling to external dissipative phenomena) sources are often considered.  Given the unresolved nature of dephasing mechanisms at low temperature, and the relevance of these issues to fundamental questions about 2D metallic phenomena, further study of these issues may prove fruitful.

\section{Acknowledgments}

This work was supported by the National Science Foundation grant NSF-DMR-9508419, FENA, as well as the German Science Foundation (DFG), within the SFB 917 (``Nanoswitches'').   We acknowledge many useful discussions with Boris Spivak, Peter Jost, Li Zhang, Ko Munakata, and George Karakonstantakis.

\end{document}